\pdfoutput=1
\documentclass[aps,onecolumn,prx,floatfix,superscriptaddress,reprint]{revtex4-1}
\usepackage[colorlinks=true, pdfstartview=FitV, linkcolor=red, citecolor=blue, urlcolor=blue]{hyperref}
\usepackage{eurosym}
\usepackage{amsbsy}
\usepackage{latexsym,epsfig,graphicx}
\usepackage{dcolumn}
\usepackage{graphicx}
\usepackage{subfigure}
\usepackage{comment}
\usepackage{color}
\usepackage{bm}
\usepackage{mathrsfs}
\usepackage{amssymb} 
\usepackage{amsthm}
\usepackage{amsfonts}
\usepackage{amsmath}
\usepackage{xspace}
\usepackage{epstopdf}
\usepackage{tabularx}
\usepackage{longtable}
\usepackage{cancel}
\usepackage[normalem]{ulem}
\usepackage[version=4]{mhchem}
\usepackage{longtable}
\usepackage{ulem}
\usepackage{color}

\begin{document}   
\title{Anomalous second-order skin modes in Floquet non-Hermitian systems}
\author{Chun-Hui Liu}
\affiliation{Department of Physics, The University of Texas at Dallas, Richardson, Texas 75080-3021, USA}
\author{Haiping Hu}
\affiliation{Beijing National Laboratory for Condensed Matter Physics, Institute of Physics, Chinese Academy of Sciences, Beijing 100190, China}
\affiliation{School of Physical Sciences, University of Chinese Academy of Sciences, Beijing 100049, China}
\author{Shu Chen}
\affiliation{Beijing National Laboratory for Condensed Matter Physics, Institute of Physics, Chinese Academy of Sciences, Beijing 100190, China}
\affiliation{School of Physical Sciences, University of Chinese Academy of Sciences, Beijing 100049, China}
\affiliation{Yangtze River Delta Physics Research Center, Liyang, Jiangsu 213300, China}
\author{Xiong-Jun Liu}
\email{xiongjunliu@pku.edu.cn}
\affiliation{International Center for Quantum Materials and School of Physics, Peking University, Beijing 100871, China}
	\affiliation{Hefei National Laboratory, Hefei 230088, China}
	\affiliation{International Quantum Academy, Shenzhen 518048, China}	
	\affiliation{CAS Center for Excellence in Topological Quantum Computation, University of Chinese Academy of Sciences, Beijing 100190, China}
\begin{abstract}
The non-Hermitian skin effect under open boundary conditions is widely believed to originate from the intrinsic spectral topology under periodic boundary conditions. If the eigenspectra under periodic boundary conditions have no spectral windings (e.g., piecewise arcs) or a finite area on the complex plane, there will be no non-Hermitian skin effect with open boundaries. In this article, we demonstrate another scenario beyond this perception by introducing a two-dimensional periodically driven model. The effective Floquet Hamiltonian lacks intrinsic spectral topology and is proportional to the identity matrix (representing a single point on the complex plane) under periodic boundary conditions. Yet, the Floquet Hamiltonian exhibits a second-order skin effect that is robust against perturbations and disorder under open boundary conditions. We further reveal the dynamical origin of these second-order skin modes and illustrate that they are characterized by a dynamical topological invariant of the full time-evolution operator.
\end{abstract}
\maketitle
\section{Introduction}
Non-Hermitian physics has recently garnered significant research interest. Many classical optical and mechanical systems, electric circuits, open quantum systems, and single-particle Green functions with non-zero self-energy are described by non-Hermitian Hamiltonians or matrices \cite{AshidaGongUeda,BBK2021RevModPhys}. Several unique features of non-Hermitian systems have been discovered, including PT symmetry \cite{PT1998Bender,PT2002Bender,PT2009Longhi,PT2011Hu,PT2012Regensburger,PT2014Feng,PT2014Zhu,PT2015Yuce,PT2017Ashida,PT2018ElGanniny}, the non-Hermitian skin effect \cite{AnomalousESLee,YaoWang,BiorthogonalKEBB,SSHLieu,ExactSolution,wang2023symmetric}, and exceptional points \cite{EP2018Shen,EP2015zhen,EP2016Doppler,EP2016xu,EP2017kozii,EP2017xu,EP2019Fu,EP2019miri,HuEP2022,HuEPclass,EPreview}. These unique features lead to a variety of tantalizing effects and phenomena in non-Hermitian systems without Hermitian counterparts \cite{SongYaoWang,HelicalDamping,InformationConstraint,Huknot,HuNHWSM,Hubartlett,NCNHWSM,SFL1,SFL2,SFL3}. Non-Hermitian systems have much richer topological phases \cite{Top2018Gong,ReflectionLJC}, as exemplified by the 38-fold symmetry classes for time-independent point gap topology \cite{KawabataShiozakiUedaSato,ZhouLee}, and 54-fold symmetry classes for time-independent line gap topology and time-dependent systems \cite{DefectsLiuChen,FloquetLHC}.

In previous studies, the paradigmatic non-Hermitian skin effect has been attributed to the spectral topology under periodic boundary conditions. Specifically, in the complex-energy plane, the appearance of skin modes with open boundaries depends on the existence of spectral winding \cite{SkinTopoOKSS,SkinTopoZYF} or finite spectral area \cite{Fang2D} under periodic boundary conditions. The skin modes can be categorized into different orders, where the $n$th-order ($n\leq d$) skin effect in $d$-dimensional systems with $L^d$ lattice sites indicates that the number of accumulated eigenstates at the $(d-n)$-dimensional boundary scales as $O(L^{d-n+1})$ \cite{SO2020Kawabata,SO2020Okugawa,SO2021Zou}. The appearance of skin modes necessitates the introduction of a generalized Brillouin zone, through which the usual bulk-edge correspondence can be restored.
And it is widely believed that the non-Hermitian skin effect (NHSE) is original from spectral winding. 

In the Floquet Hermitian system, some papers
have illustrated that topologically protected 
edge states can exist, even when the periodic boundary condition 
Floquet Hamiltonian is the identity matrix or indicates no topological 
edge modes \cite{Rudner,RoyHarper,Kitagawa}. The topological 
Floquet systems are called anomalous Floquet topological 
insulator (AFTI), which is beyond the theory 
of static topological insulators. 
Inspired by the AFTI, 
we propose a question: Is there any Floquet system where
 second-order NHSE can occur, even if 
the periodic boundary condition Floquet Hamiltonian is 
an identity matrix or indicates no second-order skin effect? 
We call it the anomalous Floquet second-order skin effect (AFSSE).

A rigorous theorem is the following: There is no NHSE for a 
finite Hilbert space Floquet Hermitian system. 
It means that some attempts based on a finite Hilbert space
Floquet Hermitian Hamiltonian cannot get AFSSE \cite{Mixed,Bessho}, although the edge-effective model is 
the Hatano-Nelson model. 
To get the AFSSE, we need to introduce non-Hermitian terms. 
The topological classification of a Floquet non-Hermitian systems
has been presented in Ref. \cite{FloquetLHC}. The non-zero topological
number
 in Table I of Ref. \cite{FloquetLHC} indicates that there can be 
 AFTI phases in the correspondence symmetry class and dimensions. 
 In the two-dimensional AFTI phases, if the adjacent boundaries'
 effective Floquet operators both have 
NHSE and are localized at the same corner, 
then the corner coupling of the adjacent boundaries does not break the NHSE.
This idea provides a method to construct the AFSSE.

This article presents a scheme that transcends the previous perception. As proof of principle, we introduce a two-dimensional Floquet driving system whose Floquet Hamiltonian is proportional to the identity matrix and has no intrinsic spectral topology. However, the Floquet Hamiltonian exhibits the second-order non-Hermitian skin effect with open boundaries. Furthermore, we demonstrate the robustness of these skin modes against perturbations and disorders and delve into their dynamical origins by explicitly working out the edge theory. We show that the skin modes are characterized by a dynamical topological invariant from the full time-evolution operator.

\section{Model}
\begin{figure}[t]
  \centerline{\includegraphics[width=3.36in]{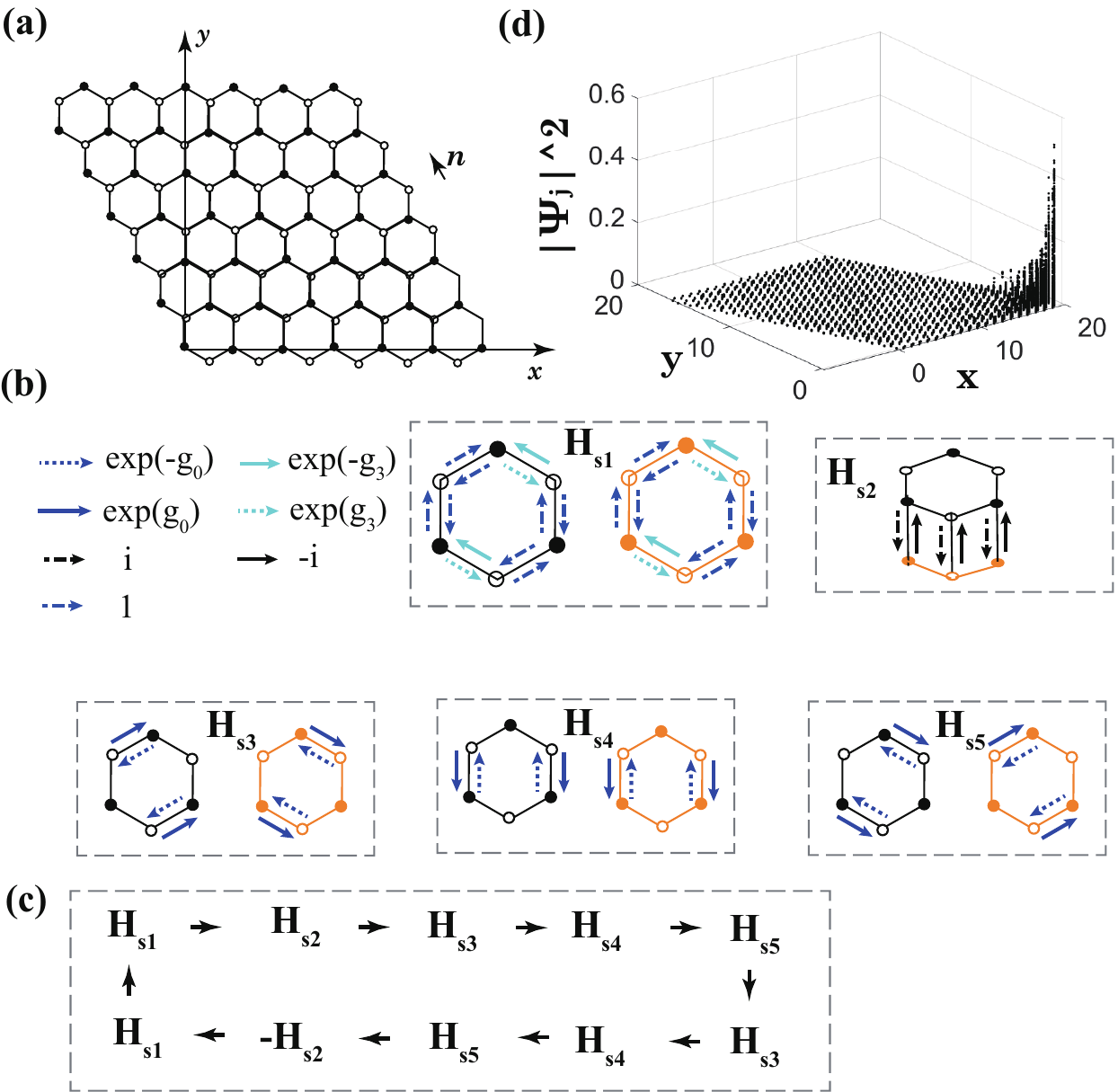}}
  \caption{Schematics of the model (top view) and driving protocol. (a) Honeycomb lattice with A (filled circles) and B (open circles) sublattices. (b) Non-Hermitian Hamiltonian $H_{sj}, (j=1,2,...,5)$ with non-reciprocal hoppings (marked by different lines) between neighboring sites. Black bonds and sites represent the top layer and the orange bonds and sites represent the bottom layer. (c) The 10-step driving sequence starting from $H_{s1}$. (d) Spatial profiles of the eigenstates. The parameters are $g_0=0.2$ and $t_1=0$. The system size is $L\times L=20\times 20$.\label{Fig1}}
\end{figure}

We consider a bilayer hexagonal lattice structure [see Fig. \ref{Fig1} (a)] and a 10-step Floquet driving sequence $H_{s1}\rightarrow H_{s2}\rightarrow ... \rightarrow H_{s10}$. The $n$-th time step inside a driving period is governed by the time-independent Hamiltonian $H_{sn}$. We set $H_{s6}=H_{s3},~H_{s7}=H_{s4},~H_{s8}=H_{s5},~H_{s9}=-H_{s2},~H_{s10}=H_{s1}$. They are given by
 \begin{equation*}
      H_{s1}({\bf k})=\left[ \begin{array}{cc}
        H_A& 0\\
     0 & H_A
      \end{array}
      \right ];       H_{s2}({\bf k})=\left[ \begin{array}{cc}
      0& -i\mathbb{I}\\
      i\mathbb{I}& 0
    \end{array}
    \right ];
\end{equation*}
\begin{equation*}
  H_{s3}({\bf k})=\left[ \begin{array}{cc}
    H_1({\bf k},g_0)& 0\\
 0 & H_3({\bf k},g_0)
  \end{array}
  \right ]; \\  
\end{equation*}
\begin{equation}
  \begin{split}
  H_{s4}({\bf k})&=\left[ \begin{array}{cc}
    H_2({\bf k},g_0) & 0\\
    0 & H_2({\bf k},g_0)
    \end{array}
    \right ];\\
  H_{s5}({\bf k})&=\left[ \begin{array}{cc}
    H_3({\bf k},g_0) & 0\\
     0 & H_1({\bf k},g_0)
      \end{array}
    \right ],  \\
    \label{Hjk4}
  \end{split}
  \end{equation}
where
 \begin{equation}
    \begin{split}
  H_1({\bf k},g)&= \left[ \begin{array}{cc}
  0 &  e^{g+i{\bf k}\cdot {\bf a}_1}\\
   e^{-g-i{\bf k}\cdot {\bf a}_1} & 0
  \end{array}
  \right ];  \\
  H_2({\bf k},g)&= \left[ \begin{array}{cc}
    0 &  e^{-g+i{\bf k}\cdot {\bf a}_2}\\
     e^{g-i{\bf k}\cdot {\bf a}_2} & 0
    \end{array}
    \right ];\\
  H_3({\bf k},g)&=\left[ \begin{array}{cc}
      0 & e^{-g+i{\bf k}\cdot {\bf a}_3}\\
       e^{g-i{\bf k}\cdot {\bf a}_3} & 0
      \end{array}
    \right ];\\
  \end{split}
\end{equation}

\begin{equation}
    H_A=H_1({\bf k},g_1)+H_2({\bf k},g_2)+H_3({\bf k},g_3).
  \end{equation}
 Here $g_0, g_1, g_2,$ $g_3$ are tunable non-reciprocal parameters between neighboring lattice sites.
  ${\bf a}_1=(-\frac{a}{2},-\frac{a}{2\sqrt{3}})$, ${\bf a}_2=(0 ,\frac{a}{\sqrt{3}})$, and
  ${\bf a}_3=(\frac{a}{2},-\frac{a}{2\sqrt{3}})$ ($a=1$ is the lattice constant). ${\bf k}$ is the lattice momentum and  $\mathbb{I}$ is the identity matrix. The time duration for each step is denoted as $t_1$, $t_2$,..., $t_{10}$, respectively. In this paper, we set $t_{j}=t_{11-j}$ for $j=1, 2, 3, 4, 5$ and $g_1=g_2=0$. Figures \ref{Fig1}(b) and \ref{Fig1}(c) sketch the non-reciprocal Hamiltonians in each step and the Floquet driving protocol.

The above driven system fulfills a type-K symmetry in the generalized Bernard-LeClair (GBL) class,
  \begin{equation}
    H({\bf k},t)=KH^*(-{\bf k},- t)K^{-1}, ~~KK^*=- \mathbb{I}, \label{Ksys}
  \end{equation}
set by the driving protocol $H_{sj}({\bf k},t)=KH_{s11-j}^*(-{\bf k},- t)K^{-1}$ ($j=1,2,3,4,5$). Here  $K=\sigma_y\otimes \mathbb{I}$ ($\sigma_y$ is Pauli matrix). The Floquet operator ($U$) is defined as the time evolution operator in one full period $T=\sum_{j=1}^{10}t_j$,
\begin{equation}
  U=e^{-iH_{s10}t_{10}}e^{-iH_{s9}t_{9}}... e^{-iH_{s1}t_{1}}.
\end{equation}
The Floquet Hamiltonian is the effective static Hamiltonian that stroboscopically tracks the time evolution. It is defined as $H_F=\frac{i}{T}ln(U)$. We always take $t_2=t_3=t_4=t_5=\frac{\pi}{2}$ in this article unless otherwise stated.

Notably, when $t_1=0$, the time-evolution operator is easily solvable and given simply by $U({\bf k})=-\mathbb{I}$, with the corresponding bulk Floquet Hamiltonian being $H_{F}({\bf k})=\frac{\pi}{T}\mathbb{I}$. Surprisingly, there are $O(L)$ eigenstates located at one single corner under the open boundary condition, while the other eigenstates are extended across the whole system, as shown in Fig. \ref{Fig1}(d) for $g_3= 0.2$. These localized states are known as the second-order skin modes and have previously been associated with the eigenspectra of a finite area on the complex plane or topological properties \cite{SO2020Kawabata,SO2020Okugawa,SO2021Zou} of the bulk Hamiltonian under periodic boundary conditions. However, in this model, the spectra of the Floquet Hamiltonian represent only a single point on the complex plane, without any topological structure or spectral area. Therefore, this model lacks any static counterparts and a complete understanding of the appearance of these skin modes requires an account of the full-time evolution.
\begin{figure}[t]
    \centerline{\includegraphics[width=3.35in]{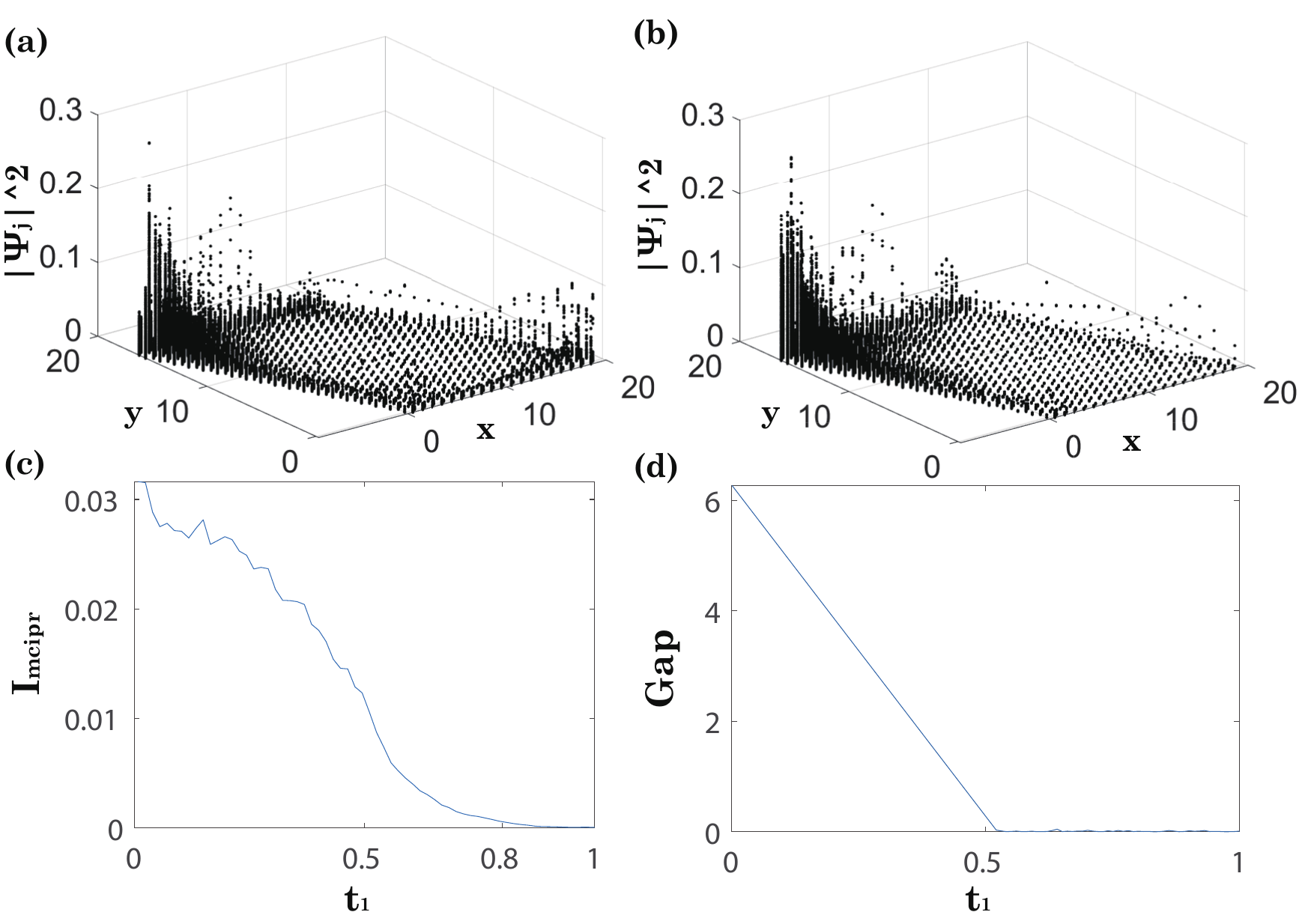}}
    \caption{Stability of second-order skin modes against perturbations. Spatial profiles of all eigenstates $|\psi_{j}({\bf r},1)|^2$ and $|\psi_{j}({\bf r},2)|^2$ for (a) $t_1=0.8$ and (b) $t_1=1$. (c) MCIPR as a function  of $t_1$. (d) Bulk spectral gap as a function of $t_1$. The open boundary is taken along $\bf{a}_3$ and its perpendicular direction. The non-Bloch band theory is used to obtain the bulk spectra. The system size is $L\times L=20\times 20$. $g_0=0.2$ and $g_3=-0.4$
      \label{Fig2}}
\end{figure}

 \section{Perturbation and disorder}
Next, we demonstrate the robustness of the second-order skin effect against perturbations (i.e., deviating from the ideal case described above.) We introduce a nonzero value for parameter $t_1$ and investigate the interplay between the first- and second-order skin effects. Additionally, we show that the second-order skin effect is resistant to disorder: the skin modes survive even in the presence of significant disorder strength.
 \subsection{Perturbation}
Let us deviate from the ideal case of Fig. \ref{Fig1} and take $g_3=-0.4$, $t_1=0.8$ with other parameters unchanged. Figure \ref{Fig2}(a) plots the spatial profiles of the eigenstates $|\psi_{j}({\bf r},1)|$ and $|\psi_{j}({\bf r},2)|$ $(j=1,2,...,4L^2-4)$ with open boundary condition. Here $z=1$, or $2$ denotes the top or bottom layer, $j$ labels the eigenstate, and ${\bf r}=(x,y)$ labels the lattice site. It is clear that there are $O(L)$ eigenvectors located at the bottom right corner and $O(L^2)$ eigenvectors located at the top left corner. That is, the first-order skin modes start to emerge, while the second-order skin modes persist. When $t_1$ increases to $1$, all eigenvectors are located at the top left corner, as depicted in Fig. \ref{Fig2}(b).

To scrutinize the second-order skin effect, we define a partial inverse partition ratio of eigenvector $\psi_{j}({\bf r},z)$
\begin{equation}
  \begin{split}
  I_{cipr}(j)= \frac{\sum_{{\bf r \in \mathcal{A}}}[|\psi_{j}({\bf r},1)|^4+|\psi_{j}({\bf r},2)|^4]}{\sum_{{\bf r}}[|\psi_{j}({\bf r},1)|^2+|\psi_{j}({\bf r},2)|^2]}.
\end{split}
\end{equation}
It has support on the bottom right part of the lattice sites with $\mathcal{A}= \left\{(x,y)|(x+\sqrt{3}y)>\frac{\sqrt{3}(L+1)}{2}, y<\frac{\sqrt{3}(L-1)}{4}\right\}$. We dub it as the corner inverse partition ratio (CIPR) and define the mean CIPR (MCIPR) over all eigenstates as
\begin{equation}
I_{mcipr}=\frac{\sum_{j}I_{cipr}(j)}{4L^2-4}.
\end{equation}
In Fig. \ref{Fig2}(c), the quantity $I_{mcipr}$ is plotted as a function of $t_1$. The decrease in $I_{mcipr}$ with increasing $t_1$ is due to the emergence of first-order skin modes and the growth of the localization length for the second-order skin modes. The small jittering in the plot may be due to finite-size effects or
the in-cell wave functions' relative changes with the variation of $t_1$. In a broad range of values for $t_1$, both first-order and second-order skin modes can be found, localized on different corners. However, as $t_1$ increases beyond a certain threshold, the first-order skin effect dominates, suppressing the second-order skin modes and pulling them towards the top left corner, as illustrated in Fig. \ref{Fig2}(b). $I_{mcipr}$ can be used as an order parameter, and the threshold value is reached at around $t_1=0.9$, where all second-order skin modes have been eliminated (i.e., they have infinite localization length). To estimate the threshold value of $I_{mcipr}$, the presence of a few extended states and the rest being localized first-order skin modes is assumed. The critical value of $I_{mcipr}$ scales as $I_{mcipr}\propto 1/(16L^2)$, where $L$ is the system size. 
In Fig. \ref{Fig2}(d), the Floquet Hamiltonian real gap at $0$ of the bulk states \cite{FloquetLHC} is plotted as a function of $t_1$. 
The Floquet Hamiltonian real gap is defined as the real line gap of $H_F$. Here, we take open boundary conditions along $\bf{a}_3$ and its perpendicular direction and utilize the non-Bloch band theory (or generalized Brillouin zone) to obtain the bulk spectra. The gap decreases and closes around $t_1=0.5$, indicating a bulk topological transition that will be discussed later.

Consider directions ${\bf a}_3$ and 
${\bf b}_1=(\frac{a}{2\sqrt{3}},\frac{a}{2})$, there is 
no NHSE if we take the open boundary condition (OBC) for the ${\bf b}_1$ direction and
periodic boundary condition (PBC) for the ${\bf a}_3$ direction. 
It means that the bulk spectrum is not dependent on
the boundary conditions in the ${\bf b}_1$ direction for $L\rightarrow \infty$. 
The bulk spectrum of the open ${\bf b}_1$ and ${\bf a}_3$ directions
is equivalent to the bulk spectrum of the open ${\bf a}_3$ direction and 
PBC on the ${\bf b}_1$ direction. Taking the Fourier transformation for 
the ${\bf b}_1$ direction, the bulk spectrum of the open ${\bf a}_3$ direction and 
PBC on the ${\bf b}_1$ direction transforms into a one-dimensional problem. 
Thus, we can use the one-dimensional non-Bloch band theory to get the spectrum.
It is similar to case A in Ref. \cite{tdnonBloch}, which also 
does not have NHSE in one direction.

If $t_1=0$, $H_F({\bf k})=\frac{\pi}{T}\mathbb{I}$. If $t_1\ne 0$, $H_F({\bf k})=H_{s1}({\bf k})+\frac{\pi}{T}\mathbb{I}$. 
In Ref. \cite{SO2020Kawabata}, Kawabata {\it et al.} found that 
there is a topological number defined by the Hamiltonian under PBC
that has correspondence with second-order skin modes.
After Fourier transforms $H_F({\bf k})$ to real space, there is no second-order skin effect in 
both the $t_1=0$ and $t_1\ne 0$ regions under OBC. 
Thus, there is no topological number defined by $H_F({\bf k})$ that
has correspondence with second-order skin modes for our model.

In Ref. \cite{Mixed}, they study the Hermitian 
Rudner-Lindner-Berg-Levin (RLBL) model. They 
found that the edge-effective Floquet operator of the RLBL model is the Hatano-Nelson model.
After cutting a hopping bound of the Hatano-Nelson model, the NHSE occurs 
for the Hatano-Nelson model. 
The significant point is that there is no NHSE in the two-dimensional 
cut hopping bound RLBL mod el. The cutting bound and
getting edge effective Floquet operator 
is not commutable.
 A rigorous theorem prohibits the existence of NHSE
 for the Floquet operator of the finite Hilbert space Hermitian system: 
 There is no NHSE for the
Floquet operator of finite Hilbert space Hermitian systems.
The eigenstates of the finite Hilbert space unitary operator $U$ 
are the same as the eigenstates of $\ln(U)$. $\ln(U)$
is a finite Hilbert space Hermitian operator. 
According to the theorem, 
for a Hermitian operator with finite Hilbert space,
 its eigenstates are orthogonal to each other, thus the eigenstates of the
finite Hilbert space unitary operator
are also orthogonal to each other.
The Floquet operator of the finite Hilbert space Hermitian system is a
finite Hilbert space unitary operator, 
its eigenstates are also orthogonal to each other.
If the Floquet operator of the finite Hilbert space Hermitian system
has NHSE, there will be $\mathcal{O}(L^{d_1})$ 
eigenstates located at $d_2$-dimensional boundary with 
$d_1> d_2$, which means that the eigenstates  
cannot be orthogonal to each other. 
It is not consistent with the fact that the eigenstates of the
Floquet operator of the finite Hilbert space Hermitian system
are orthogonal to each other. 
Thus, we get that there is no NHSE for the 
Floquet operator of finite Hilbert space Hermitian systems. 
The models discussed in Refs. \cite{Mixed} and \cite{Bessho} both 
belong to finite Hilbert space Hermitian Floquet systems. 
Thus, there is no NHSE in their models, although 
their models' edge-effective 
Floquet operator is the Hatano-Nelson model. 
Our models belong to non-Hermitian Floquet systems, which is beyond the control of this theorem. That is a 
the reason why our model can have NHSE.

Another method mentioned in Ref. \cite{SO2020Kawabata}
 to understand the second-order skin modes is to relate 
 second-order skin modes with the spectral of the edge states under 
 OBC in one direction and the PBC in another direction. If there is 
 a loop in PBC's (only one direction) edge spectra, the second-order skin effect may exist.
 This method is not rigorous and works well in many cases. 
 For the model in Sec. IV.A of Ref. \cite{FloquetLHC}, 
 there is a loop in the PBC's (only PBC in one direction) edge spectra, and there is no second-order skin effect.
 In Sec. IV, we give the effective edge theory of this model for 
$t_1=0$ (the exactly solvable limit). The effective edge theory gives an 
intuitive understanding of the second-order skin modes.
 For $t_1\ne 0$, we give a topological invariant which is defined by the full time 
 evolution operator protecting the second-order skin modes. 
The topological invariant reveals that the second-order skin effect is the origin
from anomalous Floquet topology, which may 
exist even for the PBC's Floquet Hamiltonian that is proportional to the
identity matrix. This origin leads to that ususal 
 understanding of the second-order skin effect in time-independent systems 
 cannot be used to understand this model's second-order skin effect 
 (there is no anomalous Floquet topology for the time-independent system).

    \begin{figure}[t]
      \centerline{\includegraphics[width=3.35in]{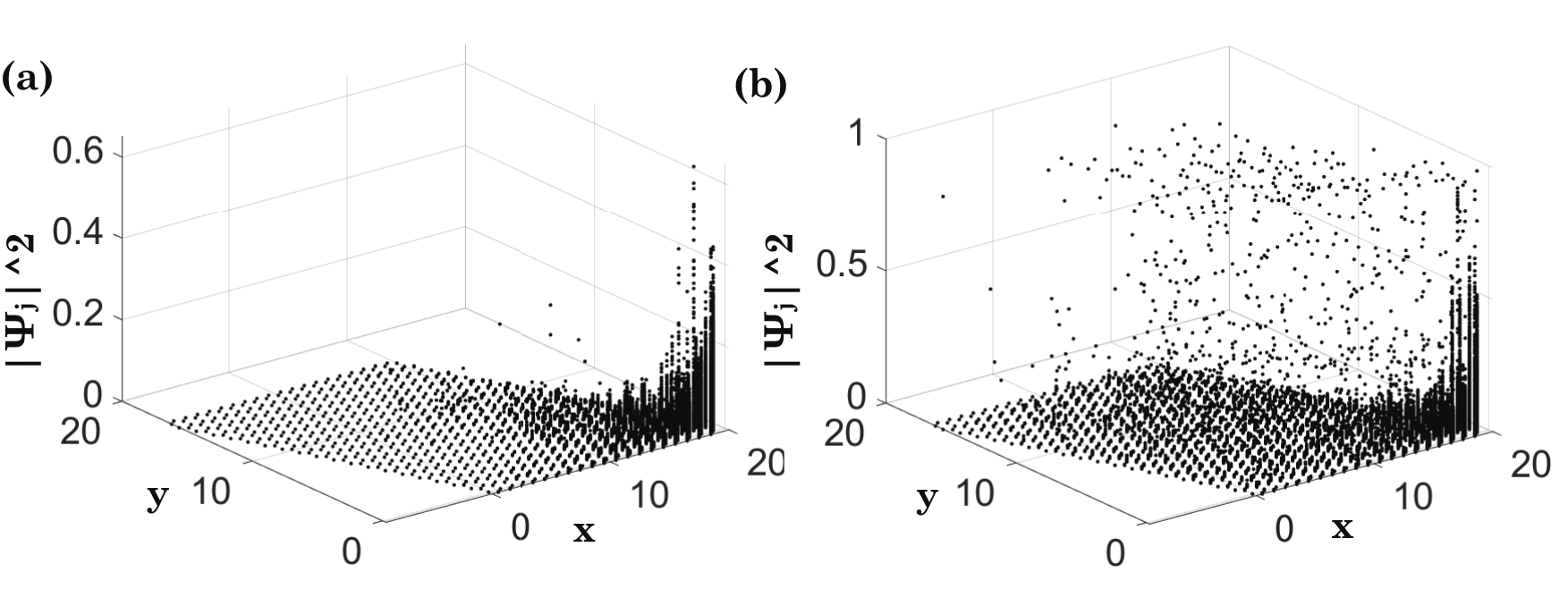}}
      \caption{Robustness of second-order skin modes with respect to disorder. Spatial profiles of all eigenstates $|\psi_{j}({\bf r},1)|^2$ and $|\psi_{j}({\bf r},2)|^2$ for disorder strength (a) $W_D=1$ and (b) $W_D=100$. Other parameters are $t_1=\pi/20$, $g_0=0.2$, and $g_3=0$. The system size is $L\times L=20\times 20$.
        \label{Fig3}}
      \end{figure}
\subsection{Disorder}
Now let us consider the effect of disorder and take $t_1=\pi/20$, $g_0=0.2$, and $g_3=0$. We add a time-independent disorder $H_{diso}$ to each step of the system's Hamiltonian:
\begin{equation}
 H_{diso}=\sum_{\bf r}[ w_d({\bf r})a_{{\bf r},1}^{\dagger}a_{{\bf r},1}+w_d({\bf r})a_{{\bf r},2}^{\dagger}a_{{\bf r},2}], \label{diso}
\end{equation}
where $a_{{\bf r},z}^{\dagger}, a_{{\bf r},z} (z=1,2)$ are creation and annihilation operations on the $\bf r$ lattice site in the $z$ layer, respectively. $w_d({\bf r})$ is a random variable with constant probability distribution in the interval $[-W_D,W_D]$. $w_d({\bf r}_i)$ and $w_d({\bf r}_j)$ $({\bf r}_i\ne {\bf r}_j)$ are independent of each other.
Figures \ref{Fig3}(a) and \ref{Fig3}(b) plot the spatial profiles of $|\psi_{j}({\bf r},1)|$ and $|\psi_{j}({\bf r},2)|$ $(j=1, 2,..., 4L^2-4)$ with $W_D=1$ and $W_D=100$, respectively. We can see that the second-order skin modes persist in a broad region of disorder strength and coexist with Anderson localization states.

\section{Edge theory and topological number}
A major advantage of our model is that it is analytically solvable under certain conditions, e.g., in the absence of disorder and $t_1=0$. This would greatly help us to gain an intuitive understanding of the emergence of second-order skin modes. For this case, the bulk Floquet operator is trivial. However, we have nontrivial edge states which are dynamically induced and the Floquet operator of
the top, bottom, left, and right edges (denoted as $U_T$, $U_B$, $U_L$, and $U_R$, respectively) are
\begin{equation}
  \begin{split}
    U_L=&\sum_{n}\frac{1}{2}[e^{-4g_0}(a_{uA,n+2}^{\dagger} -a_{dA,n+2}^{\dagger})a_{uA,n}
    \\+&e^{4g_0}(a_{uA,n}^{\dagger}-a_{dA,n}^{\dagger})a_{uA,n+2}
    \\+&e^{-4g_0}(-a_{uA,n+2}^{\dagger}+a_{dA,n+2}^{\dagger})a_{dA,n}
\\+&e^{4g_0}(a_{uA,n}^{\dagger}+a_{dA,n}^{\dagger})a_{dA,n+2}]
\\-&\sum_{n}(a_{uB,n}^{\dagger}a_{uB,n}+a_{dB,n}^{\dagger}a_{dB,n}) \label{Ul};
\end{split}
\end{equation}
\begin{equation}
  \begin{split}
U_R=&\sum_{n}\frac{1}{2}[e^{4g_0}(a_{uB,n}^{\dagger}-a_{dB,n}^{\dagger})a_{uB,n+2}
\\+&e^{-4g_0}(a_{uB,n+2}^{\dagger}-a_{dB,n+2}^{\dagger})a_{uB,n}
\\-&e^{4g_0}(-a_{uB,n}^{\dagger}-a_{dB,n}^{\dagger})a_{dB,n+2}
\\+&e^{-4g_0}(a_{uB,n+2}^{\dagger}+a_{dB,n+2}^{\dagger})a_{dB,n}]
\\-&\sum_{n}(a_{uA,n}^{\dagger}a_{uA,n}+a_{dA,n}^{\dagger}a_{dA,n}) \label{Ur};
\end{split}
\end{equation}
\begin{equation}
  \begin{split}
U_T=&\sum_{x}\frac{1}{2}[e^{4g_0}(a_{uB,x+2}^{\dagger}-a_{dB,x+2}^{\dagger})a_{dB,x}
\\+&e^{-4g_0}(a_{uB,x}^{\dagger}+a_{dB,x}^{\dagger})a_{uB,x+2}
\\-&e^{4g_0}(-a_{uB,x+2}^{\dagger}-a_{dB,x+2}^{\dagger})a_{dB,x}
\\+&e^{-4g_0}(a_{uB,x}^{\dagger}+a_{dB,x}^{\dagger})a_{dB,x+2}]
\\-&\sum_{x}(a_{uA,x}^{\dagger}a_{uA,x}+a_{dA,x}^{\dagger}a_{dA,x}) \label{Ut};
\end{split}
\end{equation}
\begin{equation}
  \begin{split}
U_B=&\sum_{x}\frac{1}{2}[e^{-4g_0}(a_{uA,x}^{\dagger}-a_{dA,x}^{\dagger})a_{uA,x+2}
\\+&e^{4g_0}(a_{uA,x+2}^{\dagger}+a_{dA,x+2}^{\dagger})a_{uA,x}
\\-&e^{-4g_0}(a_{uA,x}^{\dagger}-a_{dA,x}^{\dagger})a_{dA,x+2}
\\+&e^{4g_0}(a_{uA,x+2}^{\dagger}+a_{dA,x+2}^{\dagger})a_{dA,x}]
\\-&\sum_{x}(a_{uB,x}^{\dagger}a_{uB,x}+a_{dB,x}^{\dagger}a_{dB,x}) \label{Ub}.
\end{split}
\end{equation}
Here, $a_{zm,x}^{\dagger}$ and $a_{zm,x}$ ($z=u,d$ and $m=A,B$) are the
creation and annihilation operators on the $z$ layer, $x$-th unit cell, and $m$ sublattice.
$x$ is the cell index along the ${\bf x}$ direction [see Fig. \ref{Fig1}(a)]. $u$ and
$d$ represent the $1$ and $2$ layer, respectively. Similarly, $a_{zm,n}^{\dagger}$ and $a_{zm,n}$ are the
creation and annihilation operators on the $z$ layer, $n$-th unit cell, and $m$ site, respectively.
$n$ is the cell index along the ${\bf n}$ direction [see Fig. \ref{Fig1}(a)]. Fourier transform of Eqs. (\ref{Ul})-(\ref{Ub}) brings us to the momentum space and we have
\begin{equation}
  \begin{split}
    U_L&=\sum_{k_1}A_{k_1}^{\dagger}\left[ \begin{array}{cc}
      L_A& 0\\
   0 & -\mathbb{I}
    \end{array}
    \right ]A_{k_1};  \\
U_R&=\sum_{k_1}A_{k_1}^{\dagger}\left[ \begin{array}{cc}
  -\mathbb{I}& 0\\
0 & R_B
\end{array}
\right ]A_{k_1};  \\
U_T&=\sum_{k_2}B_{k_2}^{\dagger}\left[ \begin{array}{cc}
  -\mathbb{I}& 0\\
0 & T_B
\end{array}
\right ]B_{k_2};\\
U_B&=\sum_{k_2}B_{k_2}^{\dagger}\left[ \begin{array}{cc}
 D_A & 0\\
0 & -\mathbb{I}
\end{array}
\right ]B_{k_2}.\\
\end{split}
\end{equation}
Here
\begin{equation}
  \begin{split}
    L_A&=\frac{1}{2}\left[ \begin{array}{cc}
      e^{-4g-2ik_1}+e^{4g+2ik_1}& -e^{-4g-2ik_1}+e^{4g+2ik_1}\\
      -e^{-4g-2ik_1}+e^{4g+2ik_1}& e^{-4g-2ik_1}+e^{4g+2ik_1}
    \end{array}
    \right ];  \\
R_B&=\frac{1}{2}\left[ \begin{array}{cc}
  e^{-4g-2ik_1}+e^{4g+2ik_1}& -e^{-4g-2ik_1}+e^{4g+2ik_1}\\
  -e^{-4g-2ik_1}+e^{4g+2ik_1}& e^{-4g-2ik_1}+e^{4g+2ik_1}
\end{array}
\right ];  \\
T_B&=\frac{1}{2}\left[ \begin{array}{cc}
  e^{4g-2ik_2}+e^{-4g+2ik_2}& -e^{4g-2ik_2}+e^{-4g+2ik_2}\\
  -e^{4g-2ik_2}+e^{-4g+2ik_2} & e^{4g-2ik_2}+e^{-4g+2ik_2}
\end{array}
\right ];\\
D_A&=\frac{1}{2}\left[ \begin{array}{cc}
  e^{4g-2ik_2}+e^{-4g+2ik_2}& -e^{4g-2ik_2}+e^{-4g+2ik_2}\\
  -e^{4g-2ik_2}+e^{-4g+2ik_2} & e^{4g-2ik_2}+e^{-4g+2ik_2}
\end{array}
\right ].\\
\end{split}
\end{equation}
In our notation, $A_{k_1}=[a_{uAk_1},a_{dAk_1},a_{uBk_1},a_{dBk_1}]$,
$B_{k_2}=[a_{uAk_2},a_{dAk_2},a_{uBk_2},a_{dBk_2}]$,
$a_{zmn}=\sum_{k_1}e^{ik_1n}a_{zmk_1}$, and $a_{zmx}=\sum_{k_2}e^{ik_2x}a_{zmk_2}$, $z=u,d$ and $m=A,B$.
$k_1$ and $k_2$ are the momenta along the ${\bf n}$ and ${\bf x}$ directions, respectively. The above edge Floquet operators can be considered as similar to the Hatano-Nelson model, with non-reciprocal hoppings. 
When a full open boundary is applied in both directions, the eigenvectors of $U_L$ and $U_R$ are situated along the bottom edges, while the eigenvectors of $U_T$ and $U_B$ are located on the right edges. The combination of these edge Floquet operators results in the accumulation of edge 
states at the bottom right corner of all eigenvectors.

While the second-order skin effect can be intuitively understood through these effective edge Floquet operators, the stability of these skin modes and their origin are encoded in a dynamical topological invariant, as discussed below. For convenience, we denote $U({\bf k},t)$ as the time-evolution operator of this model. According to Table I of Ref. \cite{FloquetLHC}, this model belongs to the GBL class K2a and has $\mathbb{Z}_2$ topological classification. For Floquet systems, the description of the system's topological properties needs to take into account 
its micro-motion operator $\tilde{U}=U({\bf k},t)*e^{i H_Ft}$, in which
the $*$ operator is defined in Ref. \cite{FloquetLHC}. A Hermitianization procedure is then applied and the associated Hermitian operator is defined as
\begin{equation}
  \begin{split}
    H_{\tilde{U}}&=\left [ \begin{array}{cc}
      0& \tilde{U}({\bf k},t)\\
      \tilde{U}({\bf k},t)^{\dagger}& 0
    \end{array}
    \right ].
\end{split}
\end{equation}
By treating time $t$ as another momentum, the Hamiltonian $H_{\tilde{U}}$ can be classified as belonging to the three-dimensional Hermitian CII class \cite{ChiuRev,TeoKane}. 
The derivation of the symmetries of $H_{\tilde{U}}$ is given in Ref. \cite{FloquetLHC}. 
It should be noted that whether or not $U(k,t)$ is unitary, the $H_{\tilde{U}}$ is always Hermitian and belongs to class CII.
Denoting $|\psi_{\alpha}\rangle,(\alpha=1,2,3,4)$ as the occupied eigenstates of $H_{\tilde{U}}$ with negative eigenenergies, the non-Abelian Berry connection is defined as:
\begin{equation}
\begin{split}
A^{\alpha,\beta}({\bf k},t)=\langle \psi_{\alpha}|\nabla_{\bf k}\psi_{\beta}\rangle \cdot d{\bf k}
+\langle \psi_{\alpha}|\nabla_{t}\psi_{\beta}\rangle \cdot dt.
\end{split}
\end{equation}
The desired $\mathbb{Z}_2$ topological invariant is then the second Chern-Simons topological number $W_{CS}=CS$ $mod$ $2$ with \cite{ChiuRev,TeoKane}
\begin{equation}
\begin{split}
CS=\frac{-1}{8\pi^2}\int_{BZ\times t\in[0,T]}Tr[AdA+\frac{2}{3}A^3].
\end{split}
\end{equation}
Our numerical calculations show that $W_{CS}=1$ before the gap closure in Fig. \ref{Fig2}(d), while it is not well defined after the gap closure. The Chern-Simons topological number dictates the existence of topologically protected helical edge states. The non-reciprocal couplings along the boundaries collapse them into second-order skin modes under full open boundary conditions.

A special case worth mentioning is when $t_2=0$ and $t_3=t_4=t_5=\pi/2$. In this case, the two layers are decoupled, and each layer reduces to a previously studied model (Sec. IV A of Ref. \cite{FloquetLHC}). The topological invariants of the layers are given by three-dimensional winding numbers, with values of $1$ and $-1$, respectively. These winding numbers dictate the existence of topological edge states for each layer. If we take periodic boundary conditions along the ${\bf x}$ direction and open boundary conditions along the ${\bf n}$ direction, the type-K symmetry enforces a two-fold degeneracy (for the real part) at $k=\pi$ at the top and bottom edges (see Appendix A for details). As long as the real gap of the Floquet Hamiltonians persists, this degeneracy cannot be removed. Therefore, the topological edge states are stable and protected by the Floquet Hamiltonian's real gap, even when we deviate from the special case (e.g., by tuning $t_1$ and $t_2$ to non-zero values).

\begin{figure}[t]
  \centerline{\includegraphics[width=3.35in]{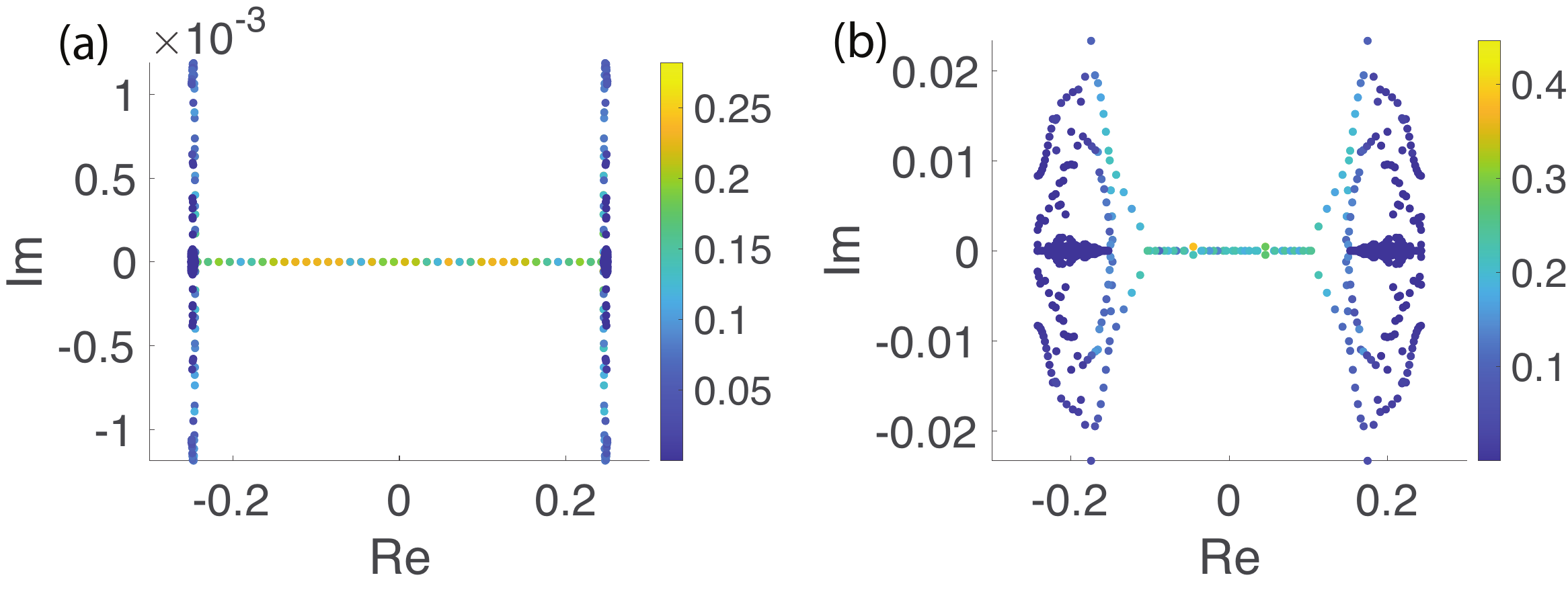}}
  \caption{OBC spectrum of $H_F$ for parameters set as $g_0=0.2$ and $g_3=-0.4$. 
  Systems size $L\times L=20\times 20$. 
  The color represents the $I_{cipr}$ of the
  corresponding eigenstate. 
  (a) $t_1=0.01$. (b) $t_1=0.2$.
    \label{Fig4}}
  \end{figure}

  Here, we demonstrate the stability of the second-order skin modes.
  The second-order skin modes cannot hybridize with 
  first-order skin modes as long as the $W_{CS}$ is non-zero. 
  Figures \ref{Fig4}(a) and \ref{Fig4}(b)
  are the spectrum of the OBC's $H_{F}$ for 
  $t_1=0.01$ and $t_1=0.2$, respectively [other parameters 
  are the same as in Fig. \ref{Fig2} (a)].
  And the color in Figs. \ref{Fig4}(a) and \ref{Fig4}(b) 
  represents the $I_{cipr}$ of the corresponding eigenstate. 
  $W_{CS}=1$ for Figs. \ref{Fig4}(a) and \ref{Fig4}(b), 
  which means that there are topologically protected
  real line gapless states for the OBC's $H_F$.
  According to Figs. \ref{Fig4}(a) and \ref{Fig4}(b), 
  we can get that the
  second-order skin modes correspond to 
  topologically protected edge states. And 
  the first-order skin modes and extended states correspond to 
  bulk states. 
   Thus, the second-order skin modes cannot hybridize with 
  first-order skin modes as long as the $W_{CS}$ is non-zero.
  
  To get a well-defined $W_{CS}$, 
  we requires the symmetry given by Eq. \ref{Ksys} and the 
  real line gap at $0$ for Floquet Hamiltonian $H_F$ \cite{FloquetLHC}. 
  As long as the perturbations are much less than the 
  real line gap of $H_F$ at $0$, the gap cannot
  close by adding the perturbations. 
  Thus, we can regard the perturbation as a weak perturbation. Otherwise,
  we can regard the perturbation as a strong perturbation. 
  On the other hand, according to the numerical results in Sec. III B,
  the second-order skin effect is 
  stable against disorder, given by Eq. (\ref{diso}). 
  The disorder given by Eq. (\ref{diso}) fulfills symmetry, given by
  Eq. (\ref{Ksys}), and almost breaks all other symmetries.
  The numerical results illustrate that the 
  second-order skin effect is stable and consistent with that 
  the second-order skin modes being protected by $W_{CS}$.

 \section{Generalization and realization}

One potential implementation of our model and driving protocol (in GBL class K2a) is through the two-dimensional unitary random walk (RW), a widely used platform of Floquet topological phases. The recipe of our setup is the anomalous Floquet topological insulator in class A \cite{RW2018chen}. By coupling two such RWs with opposite three-winding numbers ($+1$ and $-1$), we can obtain an anomalous Floquet topological insulator in class AII. To implement the non-Hermitian terms in the Floquet operators, we utilize polarization-dependent photon gain/loss, which respects type-K symmetry \cite{RW2017xiao}. The appearance of second-order skin modes in this setup should be confirmed by observing pronounced photon localizations near one of the system corners. Another possible platform for implementing our model is through open systems. A more detailed discussion of this approach can be found in Appendix B.

 \section{Conclusions}
In conclusion, we have demonstrated the emergence of second-order skin modes in a two-dimensional Floquet driving system, where the bulk Floquet band is trivially flat and 
characterized by an identity Floquet operator $-\mathbb{I}$. Our scheme surpasses prior knowledge that the non-Hermitian skin effect only arises from intrinsic bulk topology or the finite spectral area under periodic boundary conditions in the static systems. Additionally, we have shown that these second-order skin modes are 
robust against perturbations and disorders. We have further revealed the topological origin of these skin modes from the edge theory and the Chern-Simons dynamical topological invariant. Generalization to high-dimensional systems or other symmetry classes will be left for future study.

C.-H. L. thanks C.-K. Chiu for very helpful discussions, and Z.-H. Huang for helpful discussions. This work was supported by the National Key Research and Development Program of China (Grants No. 2021YFA1400900, No. 2021YFA1402104, and No. 2022YFA1405800), the National Natural Science Foundation of China (Grants No. 11825401, No. 12261160368, No.12174436, and No.T2121001), and the Strategic Priority Research Program of the Chinese Academy of Science (Grants No. XDB28000000 and No. XDB33000000). X. J. L. is also supported by the Innovation Program for Quantum Science and Technology (Grant No. 2021ZD0302000).
H. H. is also supported by the start-up grant of IOP, CAS.

\onecolumngrid
\appendix
\renewcommand{\thefigure}{S\arabic{figure}}
\renewcommand{\thetable}{S\arabic{table}}
\setcounter{equation}{0}
\setcounter{figure}{0}
\setcounter{table}{0}
\section{Proof of the real-part degeneracy of the Floquet spectra of at high symmetry point with type-K symmetry: $H({\bf k},t)=KH^*(-{\bf k},- t)K^{-1},~~KK^*=- \mathbb{I}$}
The type-K symmetry is
\begin{equation}
  H({\bf k},t)=KH^*(-{\bf k},- t)K^{-1}, ~~KK^*=- \mathbb{I}.
\end{equation}
According to Eq.(20) of Ref. \cite{FloquetLHC}, the Floquet operator fulfills
\begin{equation}
  [U^*(-{\bf k})]^{-1}=K^{-1}U({\bf k})K, ~~KK^*=- \mathbb{I}.
\end{equation}
It follows that the eigenfunction $\psi_j({\bf k})$ and eigenenergy $E_j({\bf k})$ of $U({\bf k})$ ($E_j({\bf k})=\rho_j({\bf k}) e^{i\theta_j({\bf k})}, \rho_j({\bf k})>0, 0\le \theta_j({\bf k})<2\pi$) satisfy
\begin{equation}
  U({\bf k})\psi_j({\bf k})=E_j({\bf k})\psi_j({\bf k}).
\end{equation}
Thus, we have
\begin{equation}
  \begin{split}
  &E_j({\bf k})\psi_j({\bf k})=U({\bf k})\psi_j({\bf k})=K[U^*(-{\bf k})]^{-1}K^{-1}\psi_j({\bf k}),  \\ \\
  \iff \quad &E_j({\bf k})K^{-1}\psi_j({\bf k})=[U^*(-{\bf k})]^{-1}K^{-1}\psi_j({\bf k}) ,           \\  \\
  \iff \quad &E_j({\bf k})U^*(-{\bf k})K^{-1}\psi_j({\bf k})=K^{-1}\psi_j({\bf k}),                   \\  \\
  \iff \quad &U(-{\bf k})(K^*)^{-1}\psi_j^*({\bf k})=(E_j({\bf k})^*)^{-1}(K^*)^{-1}\psi_j^*({\bf k}).
\end{split}
\end{equation}
At high symmetry point ${\bf k}_0$ (${\bf k}_0=-{\bf k}_0$), we have
\begin{equation}
  U({\bf k}_0)(K^*)^{-1}\psi_j^*({\bf k}_0)=(E_j({\bf k}_0)^*)^{-1}(K^*)^{-1}\psi_j^*({\bf k}_0).
\end{equation}
The Floquet Hamiltonian at $k_0$ is $H_F({\bf k}_0)=\frac{i}{T}ln(U({\bf k}_0))$. Here we choose the imaginary part of the $ln()$ function to be in the $[0,2\pi)$ interval. Note that both $\psi_j({\bf k}_0)$ and $(K^*)^{-1}\psi_j^*({\bf k}_0)$ are
eigenfunctions of $H_F({\bf k}_0)$, with their associated eigenenergies $\frac{1}{T}[ln[\rho_j({\bf k}_0)]i-\theta_j({\bf k}_0)]$ and $\frac{1}{T}[-ln[\rho_j({\bf k}_0)]i-\theta_j({\bf k}_0)]$. If $\psi_j({\bf k}_0)$ and $(K^*)^{-1}\psi_j^*({\bf k}_0)$ are not degenerate (with the same real part of eigenenergies), we have
\begin{equation}
  \psi_j({\bf k}_0)=e^{i\delta}(K^*)^{-1}\psi_j^*({\bf k}_0) \label{s1}.
\end{equation}
Thus
\begin{equation}
  (K^*)^{-1}\psi_j^*({\bf k}_0)= (K^*)^{-1}(e^{i\delta}(K^*)^{-1}\psi_j^*({\bf k}_0))^*=-e^{-i\delta}\psi_j({\bf k}_0),
\end{equation}
\begin{equation}
  \iff \quad \psi_j({\bf k}_0)=-e^{i\delta}(K^*)^{-1}\psi_j^*({\bf k}_0),
\end{equation}
which contradicts Eq. (\ref{s1}). Therefore, $\psi_j({\bf k}_0)$ and $(K^*)^{-1}\psi_j^*({\bf k}_0)$ are degenerate eigenfunctions (with the same real part of eigenenergies).

\section{Realization of the model in open quatum systems}
Let us consider Markovian open quantum systems, which are described by the Lindblad equation
\begin{equation}
  \frac{d\rho}{dt} 
  =\mathcal{L}[\rho]=-i[H,\rho]+\sum _{\mu}\left(2L_{\mu}\rho L_{\mu}^{\dagger}-\left\{L_{\mu}^{\dagger}L_{\mu},\rho\right\}\right) ,
  \label{lindbladeq}
  \end{equation}
  where $\rho$ is the density matrix, $H$ is the systems' Hamiltonian, and $L_{\mu}$ is the Lindblad operator. They are all time dependent. Similarly to Appendix D of Ref. \cite{InformationConstraint},
  if each Lindblad operator is a linear combination of annihilate operators, we have
 \begin{equation}
  \begin{split}
   &|\langle0| (a_{j_2}(t)a^{\dagger}_{j_1}(0)+a^{\dagger}_{j_1}(0)a_{j_2}(t))|0\rangle|^2 \\
          =&|\langle 0| a_{j_2}(t)a^{\dagger}_{j_1}(0)|0\rangle|^2  \\
          =&|\langle0|e^{\mathcal{L}(\Delta t)^{\dagger}\Delta t}[e^{\mathcal{L}(2\Delta t)^{\dagger}\Delta t}[...[
          e^{\mathcal{L}(t)^{\dagger}\Delta t}[a_{j_2}(0)]]...]]|j_1\rangle|^2  \\
          =&|Tr[e^{\mathcal{L}(\Delta t)^{\dagger}\Delta t}[e^{\mathcal{L}(2\Delta t)^{\dagger}\Delta t}[...[
            e^{\mathcal{L}(t)^{\dagger}\Delta t}[a_{j_2}(0)]]...]]|j_1\rangle \langle0|]|^2  \\
          =&|Tr[a_{j_2}(0) e^{\mathcal{L}(t)^{\dagger}\Delta t}[...[e^{\mathcal{L}(2\Delta t)^{\dagger}\Delta t}[e^{\mathcal{L}(\Delta t)^{\dagger}\Delta t}[|j_1\rangle \langle0|]]]...]]|^2  \\
          =&|Tr[a_{j_2}(0) e^{-iH_{eff}(t)\Delta t }...e^{-iH_{eff}(2\Delta t)\Delta t } e^{-iH_{eff}(\Delta t)\Delta t }|j_1\rangle \langle0|]|^2  \\
          =&|Tr[ e^{-iH_{eff}(t)\Delta t }...e^{-iH_{eff}(2\Delta t)\Delta t } e^{-iH_{eff}(\Delta t)\Delta t }|j_1\rangle \langle0|a_{j_2}(0) ]|^2  \\
          =&|\langle j_2| e^{-iH_{eff}(t)\Delta t }...e^{-iH_{eff}(2\Delta t)\Delta t } e^{-iH_{eff}(\Delta t)\Delta t }|j_1\rangle|^2.  \\
  \end{split} \label{ip2}
  \end{equation}
Here, $H_{eff}(t)=H(t)-i\sum_{\mu}^{L}L^{l\dagger}_{\mu}(t)L^l_{\mu}(t)$, and $L^l_{\mu}(t)$ is a time-dependent Lindblad operator which is a linear combination of the annihilate operator. $j_1=({\bf r}_1,z_1,m_1)$ and $j_2=({\bf r}_2,z_2,m_2)$,
 ${\bf r}_1$ and ${\bf r}_2$ are the two dimensional coordinates, $z_1,z_2=1,2$ are the layer index,
 and $m_1,m_1=A,B$ are the sublattice index. The single particle
evolution is governed by $H_{eff}(t)$. Thus, we can construct a 10-step driven open quantum system to
realize the model in the main text. The Hamiltonian and Lindblad operators in each step are
\begin{equation}
  \begin{split}
    &h_{1}=\sum_{{\bf r},z=1,2}(a^{\dagger}_{{\bf r},zA}a_{{\bf r}+{\bf a}_1,zB}
    +a^{\dagger}_{{\bf r}+{\bf a}_1,zB}a_{{\bf r},zA}+a^{\dagger}_{{\bf r},zA}a_{{\bf r}+{\bf a}_2,zB}
    +a^{\dagger}_{{\bf r}+{\bf a}_2,zB}a_{{\bf r},zA}+\beta_3 a^{\dagger}_{{\bf r},zA}a_{{\bf r}+{\bf a}_3,zB}
    +\beta_3 a^{\dagger}_{{\bf r}+{\bf a}_3,zB}a_{{\bf r},zA});  \\
    &L^1_{{\bf r},1}=\sqrt{\gamma_3}(a_{{\bf r},1A}-ia_{{\bf r}+{\bf a}_3,1B}) \\
    &L^1_{{\bf r},2}=\sqrt{\gamma_3}(a_{{\bf r},1A}-ia_{{\bf r}+{\bf a}_3,1B}) \\
    &h_{2}=\sum_{\bf r}(a^{\dagger}_{{\bf r},1A}a_{{\bf r},2A}
    +a^{\dagger}_{{\bf r},1B}a_{{\bf r},2B}+h.c.);  \\
    &h_{3}=\sum_{\bf r}\beta_0(a^{\dagger}_{{\bf r},1A}a_{{\bf r}+{\bf a}_1,1B}
    +a^{\dagger}_{{\bf r}+{\bf a}_1,1B}a_{{\bf r},1A}+a^{\dagger}_{{\bf r},2A}a_{{\bf r}+{\bf a}_3,2B}
    +a^{\dagger}_{{\bf r}+{\bf a}_3,2B}a_{{\bf r},2A});  \\
    &L^3_{{\bf r},1}=\sqrt{\gamma_0}(a_{{\bf r}+{\bf a}_1,1B}-ia_{{\bf r},1A}) \\
    &L^3_{{\bf r},2}=\sqrt{\gamma_0}(-ia_{{\bf r}+{\bf a}_3,2B}+a_{{\bf r},2A}) \\
    &h_{4}=\sum_{\bf r}\beta_0(a^{\dagger}_{{\bf r},1A}a_{{\bf r}+{\bf a}_2,1B}
    +a^{\dagger}_{{\bf r}+{\bf a}_2,1B}a_{{\bf r},1A}+a^{\dagger}_{{\bf r},2A}a_{{\bf r}+{\bf a}_2,2B}
    +a^{\dagger}_{{\bf r}+{\bf a}_2,2B}a_{{\bf r},2A});  \\
    &L^4_{{\bf r},1}=\sqrt{\gamma_0}(a_{{\bf r}+{\bf a}_2,1B}-ia_{{\bf r},1A}) \\
    &L^4_{{\bf r},2}=\sqrt{\gamma_0}(a_{{\bf r}+{\bf a}_2,2B}-ia_{{\bf r},2A}) \\
    &h_{5}=\sum_{\bf r}\beta_0(a^{\dagger}_{{\bf r},1A}a_{{\bf r}+{\bf a}_3,1B}
    +a^{\dagger}_{{\bf r}+{\bf a}_3,1B}a_{{\bf r},1A}+a^{\dagger}_{{\bf r},2A}a_{{\bf r}+{\bf a}_1,2B}
    +a^{\dagger}_{{\bf r}+{\bf a}_1,2B}a_{{\bf r},2A});  \\
    &L^5_{{\bf r},1}=\sqrt{\gamma_0}(a_{{\bf r},1A}-ia_{{\bf r}+{\bf a}_3,1B}) \\
    &L^5_{{\bf r},2}=\sqrt{\gamma_0}(a_{{\bf r}+{\bf a}_1,2B}-ia_{{\bf r},2A}) \\
    &h_6=h_3, \quad L^6_{{\bf r},1}=L^3_{{\bf r},1}, \quad L^6_{{\bf r},2}=L^3_{{\bf r},2}; \quad
    h_7=h_4, \quad L^7_{{\bf r},1}=L^4_{{\bf r},1}, \quad L^7_{{\bf r},2}=L^4_{{\bf r},2}; \\
    & h_8=h_5, \quad L^8_{{\bf r},1}=L^5_{{\bf r},1}, \quad L^8_{{\bf r},2}=L^5_{{\bf r},2}; \quad
    h_9=-h_2; \quad h_{10}=h_1, \quad L^{10}_{{\bf r},1}=L^{1}_{{\bf r},1}, \quad L^{10}_{{\bf r},2}=L^{1}_{{\bf r},2}.
  \label{lind1} 
\end{split}
\end{equation}
Here, ${\bf r}$ is the two dimensional coordinate, and $a^{\dagger}_{{\bf r},zm}$
and $a_{{\bf r},zm}$ ($z=1,2$ and $m=A,B$) are the creation and annihilation operators on
${\bf r}$ cell, $z$-th layer, and $m$ site. $\beta_0=\frac{1}{2}(e^{-g_0}+e^{g_0})$,
$\gamma_0=\frac{1}{2}(e^{g_0}-e^{-g_0})$, $\beta_3=\frac{1}{2}(e^{-g_3}+e^{g_3})$, and
$\gamma_3=\frac{1}{2}(e^{g_3}-e^{-g_3})$.
The driven sequences in one period are 
\begin{equation}
  (h_1,L^1_{{\bf r},z})\rightarrow (h_2,0)
  \rightarrow (h_3,L^3_{{\bf r},z}) \rightarrow (h_4,L^4_{{\bf r},z})
  \rightarrow (h_5,L^5_{{\bf r},z}) \rightarrow (h_6,L^6_{{\bf r},z})
  \rightarrow (h_7,L^7_{{\bf r},z}) \rightarrow (h_8,L^8_{{\bf r},z})
  \rightarrow (h_9,0) \rightarrow (h_{10},L^{10}_{{\bf r},z}) \label{seq1}.
\end{equation}
The driven time for each step is $t_1$, $t_2$, ..., $t_{10}$, respectively.
According to Eq. (\ref{ip2}), for this system we have
\begin{equation}
  \begin{split}
   &|\langle0| (a_{j_2}(T)a^{\dagger}_{j_1}(0)+a^{\dagger}_{j_1}(0)a_{j_2}(T))|0\rangle|^2 \\
          =&|\langle 0| a_{j_2}(T)a^{\dagger}_{j_1}(0)|0\rangle|^2  \\
          =&|\langle j_2| e^{-i\hat{H}_{s10}t_{10} }...e^{-i\hat{H}_{s2}t_2 } e^{-i\hat{H}_{s1}t_1 }|j_1\rangle|^2.  \\
  \end{split} \label{effe}
\end{equation}
Except for the shift by a constant matrix, $\hat{H}_{s1}$, $\hat{H}_{s2}$,..., $\hat{H}_{s1}$ take the form of $H_{s1}$, $H_{s2}$,...$H_{s10}$ in momentum space, respectively (under the basis
$A_B=[a_{{\bf k},1A},a_{{\bf k},1B},a_{{\bf k},2A},a_{{\bf k},2B}]$).
Going beyond the static limit and using the conclusions in Refs. \cite{prosen2008third,diehl2008quantum,SongYaoWang}, or Appendix A of
Ref. \cite{HelicalDamping}, the Green function
\begin{equation} 
  \Delta_{{\bf r}_1z_1m_1,{\bf r}_2z_2m_2}(t)=
  Tr[\rho(t)a^{\dagger}_{{\bf r}_1z_1m_1}a_{{\bf r}_2z_2m_2}]
\end{equation}
satisfies
\begin{equation}
  \Delta(nT)=
  e^{-2n\mathbb{I}(\gamma_3t_1+\gamma_0 (t_3+t_4+t_5))}(e^{-iH_{s10}t_{10}}...e^{-iH_{s2}t_2}e^{-iH_{s1}t_1})^n\Delta(0)
  (e^{iH_{s1}^{\dagger}t_{1}}e^{iH_{s2}^{\dagger}t_2}...e^{iH_{s10}^{\dagger}t_{10}})^n \label{Gf},
\end{equation}
where $\mathbb{I}$ is the identity matrix, $n$ is a positive integer.
According to Eq. (\ref{Gf}), the many-particle dynamics is also governed by the model in the main text.
\twocolumngrid
\bibliographystyle{apsrev4-1}
\bibliography{AFSE}
\end{document}